\documentclass[fleqn,10pt]{wlscirep}
\usepackage[utf8]{inputenc}
\usepackage[T1]{fontenc}
\usepackage{lineno}
\usepackage{booktabs}
\usepackage{comment}

\newtheorem{assumption}{Assumption}
\newcommand{\indep}{\perp \!\!\! \perp}

\title{Race and ethnicity data for first, middle, and last names}

\author[1]{Evan T. R. Rosenman}
\author[2]{Santiago Olivella}
\author[3*]{Kosuke Imai}
\affil[1]{Harvard Data Science Initiative, Harvard University}
\affil[2]{Department of Political Science, University of North Carolina at Chapel Hill}
\affil[3]{Department of Government and Department of Statistics, Harvard University}

\affil[*]{corresponding author: Kosuke Imai (imai@harvard.edu)}

\begin{abstract}
  We provide the largest compiled publicly available dictionaries of first, middle, and last names for the purpose of imputing race and ethnicity using, for example, Bayesian Improved Surname Geocoding (BISG).  The dictionaries are based on the voter files of six Southern states that collect self-reported racial data upon voter registration. Our data cover a much larger scope of names than any comparable dataset, containing roughly one million first names, 1.1 million middle names, and 1.4 million surnames. Individuals are categorized into five mutually exclusive racial and ethnic groups --- White, Black, Hispanic, Asian, and Other --- and racial/ethnic counts by name are provided for every name in each dictionary.  Counts can then be normalized row-wise or column-wise to obtain conditional probabilities of race given name or name given race. These conditional probabilities can then be deployed for imputation in a data analytic task for which ground truth racial and ethnic data is not available.
\end{abstract}

\begin{document}

\flushbottom
\maketitle

\thispagestyle{empty}

%\noindent Please note: Abbreviations should be introduced at the first mention in the main text – no abbreviations lists or tables should be included. Structure of the main text is provided below.

\section*{Background \& Summary}

In the absence of ground truth data, researchers frequently seek to predict the race and ethnicity of individuals. Such prediction tasks are essential for identifying racial disparities in areas such as housing, public health, and political participation. Because researchers often have access to information about individuals' names and addresses, many modern approaches typically utilize these data to make informed predictions about individual race. A leading methodology is Bayesian Improved Surname Geocoding (BISG), which uses Bayes' rule to obtain predictions about an individual's race by combining Census-derived race-name distributions with the decennial Census counts for certain geographical units such as Census blocks \cite{elli:etal:08,elli:etal:09,fisc:frem:06,imai2016improving}. BISG has been deployed in studies on racial disparities in policing \cite{edwa:lee:espo:19}, eviction \cite{hepb:loui:desm:20}, and voter turnout \cite{frag:18}.

The validity of BISG and similar methodologies is, however, reliant on accurate estimates of the race-name relationship, operationalized as either the conditional probabilities of race given name or the conditioal probabilities of name given race. A commonly-used BISG implementation, including the R package \texttt{WRU} \cite{wru}, combines two U.S. Census datasets to estimate these distributions for surnames. The first is the Census Bureau’s surname list, which provides the racial distribution of surnames appearing at least 100 times. The 2010 version of this list contains about 160,000 names, covering approximately 90\% of the population \cite{imai2016improving}. The second source is the Census's Spanish surname list, which contains roughly 12,000 common Hispanic surnames, approximately half of which are not in the Census surname list. These lists can still fail to contain the surnames of disproportionately large subsets of minority populations, especially Asian Americans \cite{2205.06129}. Other resources for race-name distributions \cite{tzioumis2018demographic} suffer from similar challenges in terms of coverage.

Our new dataset, available in the Harvard Dataverse \cite{DVN/SGKW0K_2022}, substantially enhances the available data for estimating race-name distributions. We make use of voter files from six states in the American South which collect data about voters' self-reported race and ethnicity. The six states are: Alabama, Florida, Georgia, Louisiana, North Carolina, and South Carolina. The voter files --- a compendium of all registered voters in a given state at a given time --- are sourced from L2, Inc., a leading national non-partisan firm and the oldest organization in the United States that supplies voter data and related technology to candidates, political parties, pollsters, and consultants for use in campaigns. We consider recent voter files in each state, comprising nearly 38 million total registered voters across the six states. More than 90\% of voters provide self-reported race information --- data that is widely accepted as a standard for ground truth \cite{tzioumis2018demographic}. These datasets allow us to compile rich dictionaries mapping first, middle, and last names to their empirical frequencies by race: the first name dictionary contains just shy of one million unique names; the middle name dictionary contains roughly 1.1 million names; and the last name dictionary contains 1.4 million unique names. From these data, researchers can use row-wise or column-wise normalization to obtain conditional probabilities of name given race or conditional probabilities of race given name.

In a separate paper \cite{2205.06129}, we demonstrate the practical utility of this data enhancement, which is now incorporated into the latest version of \texttt{WRU}. We analyzed the voter file data from the aforementioned Southern states, using a leave-one-out approach to assess how the additional data improves the out-of-sample predictive accuracy while maintaining a high calibration of predicted probabilities. Using the standard BISG procedure with the additional name data, and assigning each individual to their maximum a posteriori class prediction, we achieved a misclassification rate of 13.2\%. This represented a significant improvement over the error rate of 16.7\% achieved using only the surname data sourced rom the census.  We also found that particularly substantial improvements were made for racial minorities.  This led us to believe that our data, if made publicly available, could benefit other researchers by improving the accuracy of individual race predictions.

\section*{Methods}

\subsection*{Voter Files}

Voter files for all six U.S. Southern states were sourced from L2, Inc. The voter files were dated between October 2020 and February 2021. A summary of the relevant data for each state can be found in Table~\ref{tab:raceDist}.  We find that self-reported racial data was listed on the file for the overwhelming majority of voters across states. For Florida, Georgia, and North Carolina, approximately 3.7\%, 8.5\%, and 10.5\% of registered voters do not have self-reported race.

While every Southern state is majority White, each has a large Black population, ranging from 14\% to 31\% of the total population of registered voters in the state. There are relatively fewer Hispanic voters in the South, with such voters comprising a large proportion of the population in Florida (18\%), but less than 5\% in all other states. The Asian proportion of the population is even smaller, ranging from less than 1\% of the population (Alabama) to nearly 3\% (Georgia). Even for racial groups that comprise a small proportion of the population, these files provide full name and self-reported racial data for hundreds of thousands of voters. 

\begin{table}[h]
\begin{tabular}{llrrrrrrr} \toprule
\textbf{State}   & \textbf{\begin{tabular}[c]{@{}l@{}}Date                                                                                                                                                                                                                                                                                                                                                       \\ Sourced\end{tabular}} & \multicolumn{1}{l}{\textbf{\begin{tabular}[c]{@{}l@{}}Total \# of Reg \\ Voters (Million)\end{tabular}}} & \multicolumn{1}{l}{\textbf{\% White}} & \multicolumn{1}{l}{\textbf{\% Black}} & \multicolumn{1}{l}{\textbf{\% Hispanic}} & \multicolumn{1}{l}{\textbf{\% Asian}} & \multicolumn{1}{l}{\textbf{\% Other}} & \multicolumn{1}{l}{\textbf{\% No Race}} \\ \midrule
Alabama          & 02/07/21                                                         & 3.5                                                                                                      & 69.5                                  & 26.9                                  & 1.2                                      & 0.7                                   & 1.1                                   & 0.6  \\
Florida          & 02/07/21                                                         & 14.2                                                                                                     & 60.6                                  & 13.7                                  & 17.8                                     & 2.1                                   & 2.1                                   & 3.7  \\
Georgia          & 02/07/21                                                         & 7.1                                                                                                      & 52.7                                  & 30.3                                  & 3.7                                      & 2.7                                   & 2.2                                   & 8.5  \\
Louisiana        & 01/26/21                                                         & 3.0                                                                                                      & 63.0                                  & 31.3                                  & 1.6                                      & 1.0                                   & 3.1                                   & 0.0  \\
North Carolina   & 02/07/21                                                         & 6.7                                                                                                      & 62.7                                  & 20.2                                  & 3.0                                      & 1.4                                   & 2.2                                   & 10.5 \\
  South Carolina & 10/05/20                                                         & 3.3                                                                                                      & 69.0                                  & 27.0                                  & 1.8                                      & 1.1                                   & 1.0                                   & 0.1  \\
  \midrule
  Total          &                                                                  & 37.8                                                                                                     & 61.2 &	21.8 &	8.3 &	1.8 &	2.0 &	4.9 \\
  \bottomrule           
\end{tabular}
\caption{\label{tab:raceDist} Aggregate racial distributions for voter files in each state.}
\end{table}

\subsection*{Data Processing}

Data processing for first and middle names was minimal. For each name type, name tallies by race were computed by iterating through the file.  Voters were categorized into five racial groups --- White, Black, Hispanic, Asian, and Other --- based on self-reported racial data. Each of these categories maps directly to categories reported on the L2-sourced voter file, with the exception of the Other category, which is an amalgam of the Native American and Other categories on the file. 

Note that the L2-sourced file incorporates some post-processing on top of the base voter file. While states often collect race and ethnicity data separately --- such that individuals may identify as, e.g., both Black and Hispanic --- L2 collapses these categorizations into a single field. Using a fairly standard convention \cite{tzioumis2018demographic}, Hispanic voters are grouped into a single category regardless of race, while non-Hispanic voters are identified with their self-reported racial category.

Once the name tallies were obtained, names were cast to uppercase. A small number of names that contained fully non-ASCII characters were collapsed into a NULL category that is also included in the dictionaries. While a NULL first name is quite rare, a NULL middle name is quite common, as many individuals have only a first name and a surname. The final dictionaries provide a rich detail about race-name distributions. Our first name dictionary contains roughly 993,000 unique names, while our middle name dictionary contains 1.09 million unique names and our last name dictionary contains 1.42 million unique names. These data can be normalized column-wise to generate conditional probabilities $\mathbb{P}(\text{name} \mid \text{race})$, and also row-wise, to generate conditional probabilities $\mathbb{P}(\text{race} \mid \text{name})$. These conditional probabilities can then be deployed for any race prediction task.

\subsection*{Potential Discrepancies: Voter files vs. U.S. Population}

Our goal is to provide a dataset that reasonably captures name-race distributions for the United States population. However, our data incorporates two plausible sources of bias. First, our data is drawn only from the population of registered voters. Aside from the standard restrictions that voters be U.S. citizens and over the age of 18, there are considerable demographic differences between the populations of registered voters and nonvoters \cite{berry2010voters}.

Nonetheless, under the following assumption, the conditional distribution of race given name $\mathbb{P}(\text{race} \mid \text{name})$ is identical between registered voters and nonvoters.
\edef\oldassumption{\the\numexpr\value{assumption}+1}
\setcounter{assumption}{0}
\renewcommand{\theassumption}{\oldassumption.\alph{assumption}}
\begin{assumption}\label{indepReg-a}
  Conditional on name $n_i$, the individual's race $r_i$ and voter registration status $\text{reg}_i$ are independent. That is,
$$r_i \indep \text{reg}_i \mid n_i.$$
\end{assumption}
This assumption is violated if, for example, Black and white voters with an identical surname have different propensities to register to vote.  Alternatively, our data can also be used to obtain the conditional probabilities of race given name if the following assumption holds:
\begin{assumption}\label{indepReg-b}
Conditional on the race $r_i$ of an individual $i$, the individual's name $n_i$ and voter registration status $\text{reg}_i$ are independent. That is, 
$$n_i \indep \text{reg}_i \mid r_i.$$
\end{assumption}
The assumption is violated, for example, if Black voters with different surnames have varying propensities to register.  If either Assumption~\ref{indepReg-a}~or~\ref{indepReg-b} is satisfied, researchers can reliably use our data in combination with the Census name data to obtain relevant conditional probabilities through one of the BISG formulae.

The second source of plausible bias is geographic: our data is drawn from only six states, whose demographics differ from those of the entire country. We can quantify the discrepancies between our sample and the broader American population on the measure of race and ethnicity. Among the roughly 36 million individuals whose races are provided in the voter files, approximately 64\% identify as White, 23\% as Black, 9\% as Hispanic, 2\% as Asian, and 2\% are in the ``Other'' category. While the proportion of White Americans in our sample roughly matches the national population, Hispanic and Asian Americans are severely underrepresented, and Black Americans are modestly overrepresented \cite{census:2020}.  

Like the potential selection bias due to voter registration discussed above, however, what matters is whether the conditional (rather than marginal) probabilities match those derived from the target national population. In particular, the relevant conditional probabilities remain valid so long as one of the following two assumptions holds, \edef\oldassumption{\the\numexpr\value{assumption}} \setcounter{assumption}{0} 
\begin{assumption}\label{indepGeo-a}
Conditional on name $n_i$, the individual's race $r_i$ and voter geolocation $\text{geo}_i$ are independent. That is, 
$$r_i \indep \text{geo}_i \mid n_i\,.$$
\end{assumption}
\begin{assumption}\label{indepGeo-b}
  Conditional on the race $r_i$ of an individual $i$, the individual's name $n_i$ and geolocation $\text{geo}_i$ are independent.
$$n_i \indep \text{geo}_i \mid r_i.$$
\end{assumption}
Assumption~\ref{indepGeo-a} is violated if, for instance, two Hispanic and Black voters who share the same surname are systematically inclined to live in different places. In turn, Assumption~\ref{indepGeo-b} would fail to hold if, for example, knowing the first name of two White voters could help us predict their place of residence.

Although fully verifying these assumptions is impossible, we can evaluate their plausibility by comparing conditional distributions derived from our data to those derived from other sources that have broader geographic and domain diversity.

\section*{Technical Validation}

In this section, we empirically test whether these assumptions are plausible by comparing our data against other publicly available race-name dictionaries: one based on the U.S. Census, and another based on mortgage applications in the entire United States.     

\subsection*{Comparison against Census surname list}

We first link our last name dictionary to the 2010 Census surname list and compare the quantities $\mathbb{P}(\text{race} = r \mid \text{last name})$ and $\mathbb{P}( \text{last name} \mid \text{race} = r)$ for each $r \in \{\text{White, Black, Hispanic, Asian, Other}\}$ for every one of the linked names. The 2010 version of the surname list is used because the 2020 Census surname list has not yet been released. Of the 162,253 names on the 2010 Census surname list --- representing all names that appeared more than 100 times in that census --- we are able to link over 97\% of them to entries in our last name dictionary. While the linked entries represent only 11\% of the total names in our dictionary, they cover more than 90\% of the 36 million individuals with a reported race in the voter files.

\begin{figure}[t]
    \centering
    \includegraphics[width = \textwidth]{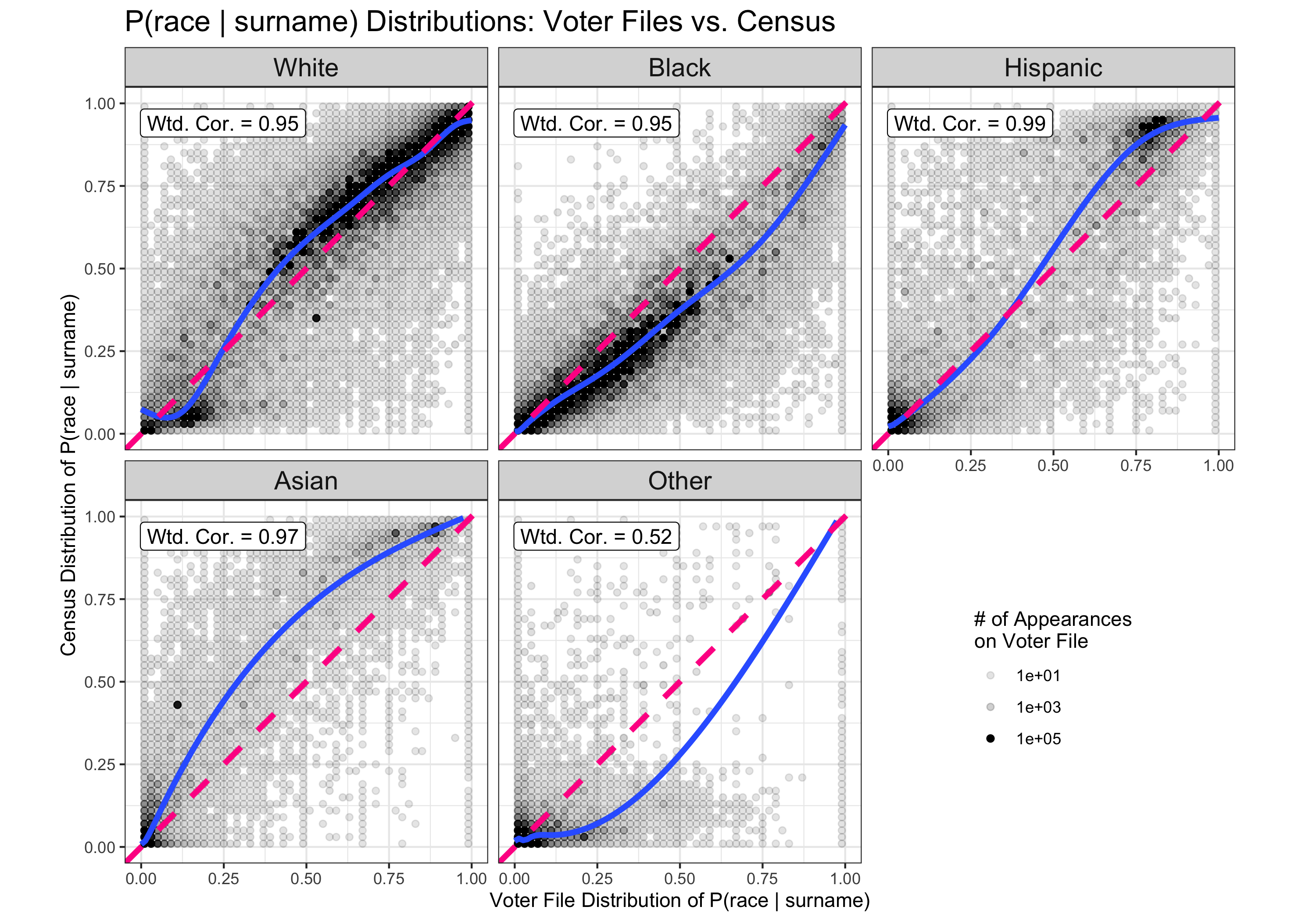}
    \caption{$\mathbb{P}(\text{race} = r \mid \text{last name})$ values for each racial and ethnic group $r$,  computed for linked names via the census ($y$ axis) and via the voter files ($x$ axis). Data is coarsened such that each dot represents a 2\% range along each axis. The dots' opacity corresponds to the total number of voter file appearances for names that fall into each range.}
    \label{fig:census_raceGivenName}
\end{figure}

\begin{figure}[t]
    \centering
    \includegraphics[width = \textwidth]{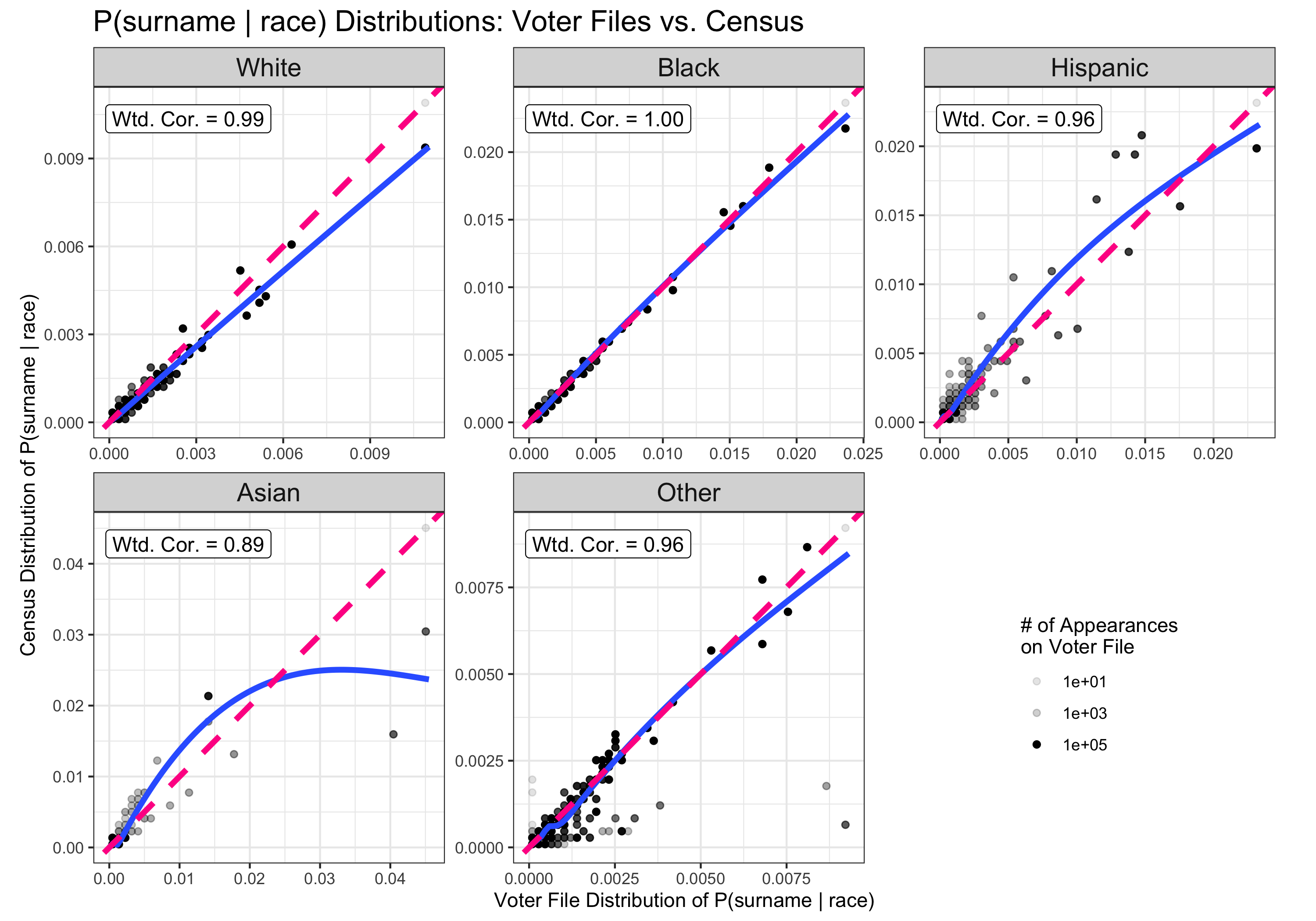}
    \caption{$\mathbb{P}(\text{race} = r \mid \text{last name})$ values for each racial and ethnic group $r$, computed for linked names via the census ($y$ axis) and via the voter files ($x$ axis). Data is coarsened such that each dot represents a 2\% range along each axis. The dots' opacity corresponds to the total number of voter file appearances for names that fall into each range.}
    \label{fig:census_nameGivenRace}
\end{figure}

Figures~\ref{fig:census_raceGivenName} and \ref{fig:census_nameGivenRace} plot the comparisons. In Figure~\ref{fig:census_raceGivenName}, the census distribution of $\mathbb{P}(\text{race} = r \mid \text{last name})$ is given on the $y$-axis, whereas the voter file distribution of the probabilities is given on the $x$-axis. The same holds for $\mathbb{P}(\text{last name} \mid \text{race} = r)$ in Figure \ref{fig:census_nameGivenRace}. For ease of visualization, we plot a coarsened version of the data, where the the opacity of each dot reflects the total number of voter file appearances of surnames that fall into that region of the plot. The weighted correlations reported on each plot are computed using all the data, with weights also corresponding to the frequency with which such names appear on the voter files. In solid blue, we plot simple smoothed trendlines, computed via a generalized additive model. The trendline is weighted by the number of appearances of a surname on the voter file. In dashed pink, we plot the 45-degree line, indicating perfect agreement between the two sets of conditional probabilities.

Figure~\ref{fig:census_raceGivenName} simultaneously assesses the plausibility of Assumptions \ref{indepReg-a} and \ref{indepGeo-a}. Weighted correlations are 95\% or higher for all racial and ethnic groups except for the ``Other'' category. This indicates that there is broad concordance between the census data and the voter file data. That is, surnames that the census finds to be strongly indicative of membership in a particular racial group are typically indicative of the same group membership in the voter file data. These high correlations provide evidence for the approximate validity of Assumptions~\ref{indepReg-a}~and~\ref{indepGeo-a}. The sole group for which strong concordance is not observed is the ``Other'' group, for which the weighted correlation falls to a much less impressive 52\%. We are not surprised by this result, as the ``Other'' category is quite small and an amalgam of several smaller racial and ethnic groups. Further discussion of these discrepancies can be found in the Appendix.

Moreover, the best fit lines (plotted in solid blue) adhere closely to the 45-degree line (in dashed pink), indicating that while there may be some deviation from the census probabilities, there is little systematic bias. Conditional on name, we see almost no evidence that White and Hispanic individuals are over- or underrepresented in the voter files relative to the Census. We do see that, conditional on name, there is evidence of slight overrepresentation of Black and Other individuals in the voter files relative to the census, and modest underrepresentation of Asian individuals relative to the Census. Because these discrepancies mirror the relative over- and under-representations of these racial groups in the South relative to the United States, these trends likely indicate minor violations of Assumption~\ref{indepGeo-a}. However, they could also partially be attributable to violations of Assumption~\ref{indepReg-a}, or to the temporal discrepancy when using 2010 Census probabilities to analyze 2020 voter files.

Figure~\ref{fig:census_nameGivenRace} compares the conditional probabilities of last name given race between our data and the 2010 Census name list, and hence simultaneously asseses the plausibility of Assumptions \ref{indepReg-b}~and~\ref{indepGeo-b}. For White, Black, Hispanic, and Other individuals, we see high correlation between the probabilities computed on the two datasets, and no evidence of systematic bias. For Asian Americans, the weighted correlation is still nearly $0.90$, but it is notably lower than the other racial groups. Two outliers are visible on the righthand side of the plot, corresponding to two popular surnames --- ``Nguyen'' and ``Patel'' --- which drive down the correlation. These two names, which are common Vietnamese and Indian surnames, respectively, are more common among Asian individuals on the voter files than in the Census. This is almost certainly due to the fact that Vietnamese-Americans and Indian-Americans comprise a larger proportion of Asian Americans in the  South than in the country at large \cite{saka:13, budiman2019key}, inducing a modest violation of Assumption~\ref{indepGeo-b}. We note that these data may not perform optimally when making predictions on Asian populations whose distribution of national origin differs markedly from that of Southern states, but we otherwise find that Assumptions~\ref{indepReg-b}~and~\ref{indepGeo-b} hold reasonably well in these data. 

This validation study does not test the plausibility of assumptions among the roughly 1.2 million names in our surname dictionary that we are unable to link to the Census. The results, however, show that for the overwhelming majority of our voter file population --- the 91\% of individuals that can be linked to the census surname list --- the required assumptions appear plausible. 

\subsection*{Comparison against the first-name dataset from mortgage applications}

We also compare our data against the database of first name racial distributions compiled by Tzioumis \cite{tzioumis2018demographic}. These data were created by merging three proprietary mortgage application datasets, drawn from applications submitted between 2007 and 2010. The lenders comply with requirements under the Home Mortgage Disclosure Act (HMDA), meaning that they collect self-reported race and ethnicity data about the applicants. The total sample size of first name observations is about 2.66 million, which is smaller than our voter file data of 36 million. After some careful processing steps, Tzioumis obtains a list of 4,250 names along with their corresponding race and ethnicity distributions.

We are interested in comparing against these data for several reasons. First, as far as we are aware, this is the sole database of first-name racial distributions that is publicly available. Second, Tzioumis directly cautions against using voter file data for the purposes of race imputation, writing:
\begin{quote}
In terms of voter registration data, the vast majority of states either do not have information on the race and ethnicity of the voters, or have restrictions on the use of the data, or both. Moreover, in terms of data integrity when combining data across states, there may be concerns as data maintenance and voter purge practices vary substantially across election boards
\end{quote}
We echo Tzioumis in noting that the states which provide race data on their voter files often have different demographics than the broader U.S. population. However, our analysis in the prior section indicates that, once we focus on conditional distributions, the magnitude of this bias shrinks considerably.

Unfortunately, given the lack of Census first and middle name lists, we cannot directly verify the validity of our first and middle name lists.  Nor can we examine the appropriateness of Tzioumis' first name list.  Thus, our comparison here is confined to the simple description of the differences between the two first-name data sets.

Of the 4,250 unique names on the mortgage application list, we are able to link all of them to entries in our first name dataset. The linked entries represent less than 0.5\% of the total names in our dataset, reflecting the fact that our dictionary is significantly larger than the insurance names dictionary. Nonetheless, the linked entries cover 84.8\% of the 36 million individuals from the voter files who report their race. This owes to the concentration of first names: the vast majority of individuals have a popular first name, while concentration is considerably lower for surnames \cite{tzioumis2018demographic}. 

The racial and ethnic categorization of the mortgage application data is quite similar to ours. In particular, self-identified Hispanics are unified into a single group, regardless of their listed race, while the White, Black, and Asian groups are defined as non-Hispanic members of those racial groups. Tzioumis provides two additional categories --- American Indian (or Alaska Native (0.17\% of observations) and Multi-race (0.16\%) --- that are not present in our categorization. For the purposes of comparison, we aggregate these two groups and compare them against our ``Other'' group (which comprises 2.1\% of our sample).

\begin{figure}[h]
    \centering
    \includegraphics[width = \textwidth]{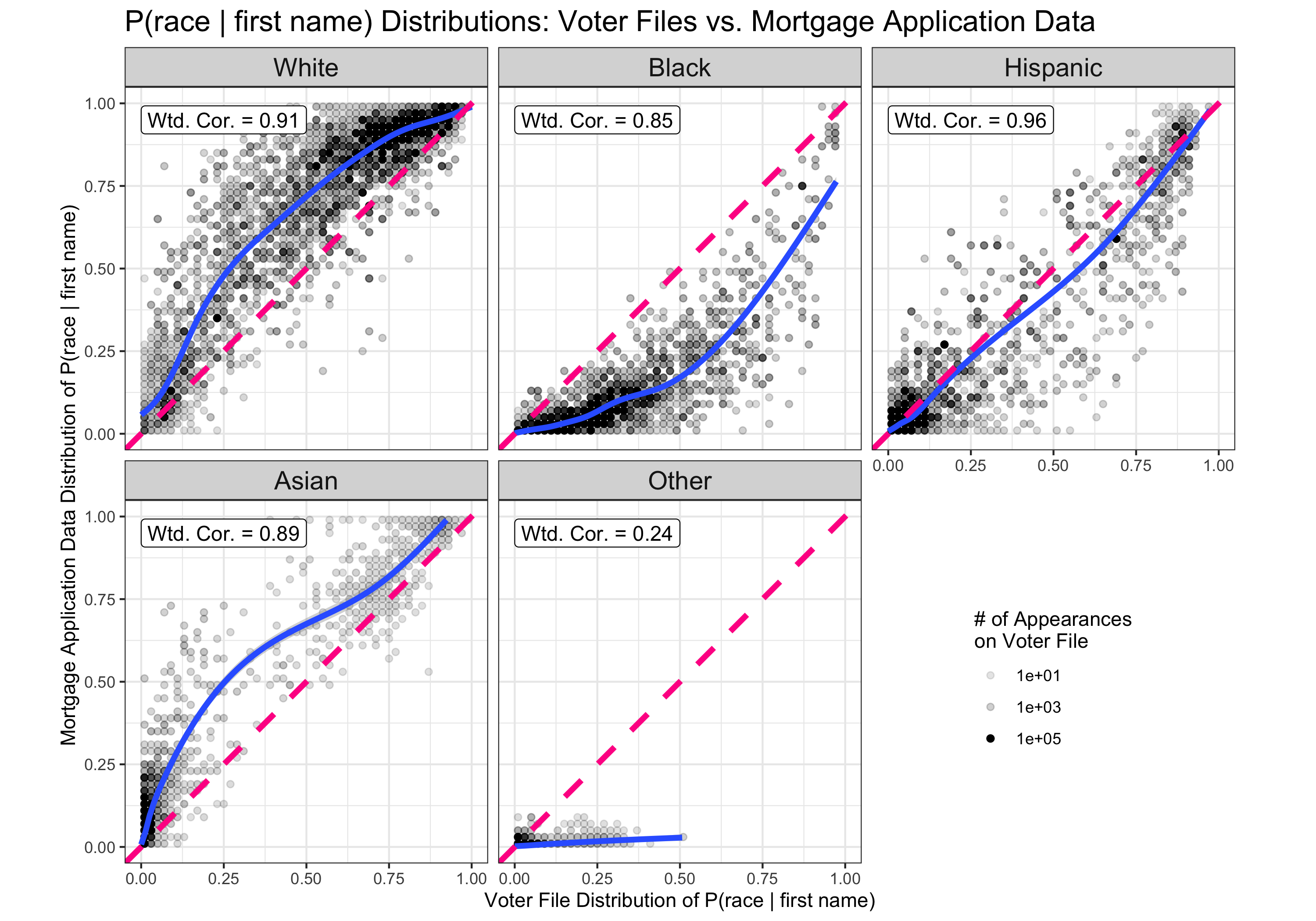}
        \caption{$\mathbb{P}(\text{race} = r \mid \text{first name})$ values for each racial and ethnic group $r$,  computed for linked names via the Tzioumis mortgage application data ($y$ axis) and via the voter files ($x$ axis). Data is coarsened such that each dot represents a 2\% range along each axis. The dots' opacity corresponds to the total number of voter file appearances for names that fall into each range.}
    \label{fig:tzioumis_raceGivenName}
\end{figure}

In Figure~\ref{fig:tzioumis_raceGivenName}, we plot the comparisons for $\mathbb{P}(\text{race} \mid \text{first name})$, using an identical set-up to the census comparison. The mortgage application distribution of $\mathbb{P}(\text{race} = r \mid \text{first name})$ is given on the $y$-axis and the voter file distribution of the probabilities is given on the $x$-axis. We again report weighted correlations and provide a smoothed trendline via a generalized additive model. 

\begin{figure}[h]
    \centering
    \includegraphics[width = \textwidth]{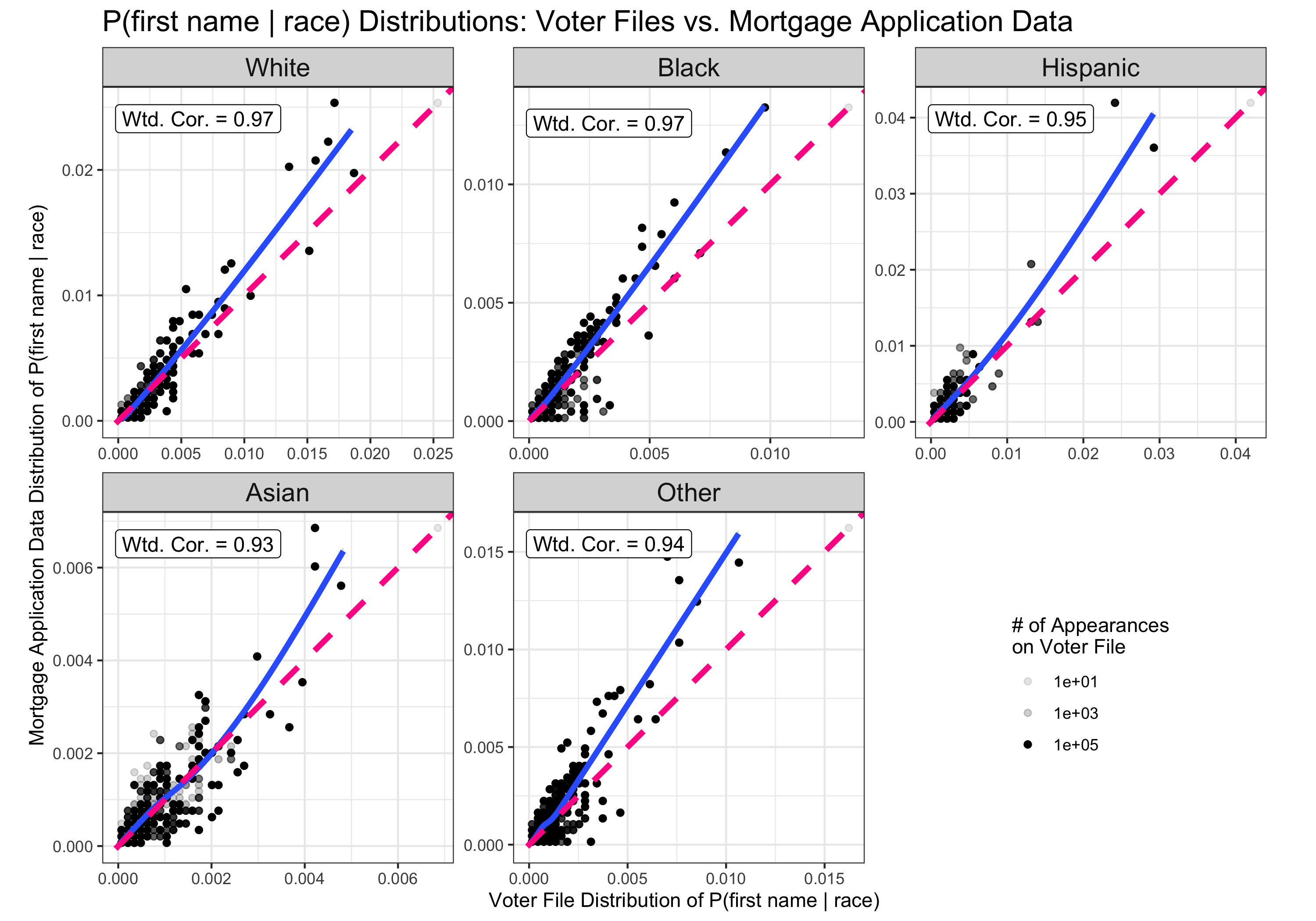}
        \caption{$\mathbb{P}(\text{first name} \mid \text{race} = r )$ values for each racial and ethnic group $r$,  computed for linked names via the Tzioumis mortgage application data ($y$ axis) and via the voter files ($x$ axis). Data is coarsened such that each dot represents 2\% of the range of each axis. The dots' opacity corresponds to the total number of voter file appearances for names that fall into each range.}
    \label{fig:tzioumis_nameGivenRace}
\end{figure}

The comparison is instructive in a number of ways. Weighted correlations are reasonably high for all groups except the ``Other'' category, though they are much lower than in comparison with the Census surname list. Moreover, there appear to be somewhat pronounced systematic discrepancies between the racial distributions conditional on the name. The mortgage application data appears to assign considerably higher probabilities of being White or Asian than the voter file data, and considerably lower probabilities of being Black. 

Aggregating the Tzioumis data, we find that their sample is 82.3\% White, 4.2\% Black, 6.9\% Hispanic, 6.3\% Asian, and 0.3\% ``Other.'' These marginals are not entirely unexpected: it is a well-established finding that White people are significantly more likely than members of any other racial or ethnic group to own their own homes \cite{goodman2018homeownership}. It is hence no surprise that mortgage applications skew disproportionately toward White applicants. 

We also recall that the voter file data does not appear to be systematically Whiter than the U.S. population, conditional on last name. Hence, we suspect that the differences in White probabilities may be driven largely by overrepresentation of White applicants in the mortgage application data. Baselining against the Census results, it appears that that our data may tend to assign slightly too high probabilities of being Black conditional on first name, while the Tzioumis data may tend to assign too low probabilities. There does not appear to be a systematic difference among Hispanic first names. The results are less clear for Asian Americans, but the Tzioumis data may provide more accurate national probabilities for this group due to better representation among mortgage applicants.

The ``Other'' categorization is, in aggregate, nearly 10 times less frequent among the mortgage application data than in the voter file data. The poor correlation exhibited between the datasets on $\mathbb{P}(\text{race} = \text{``Other''} \mid \text{first name})$ is largely driven by this discrepancy in marginal composition. 

In Figure~\ref{fig:tzioumis_nameGivenRace}, we plot the comparisons for the inverted conditional probabilities, $\mathbb{P}(\text{first name} \mid \text{race})$. All features of the plot remain the same. We observe very high correlations for all racial groups (even the ``Other'' group) for name probabilities conditional on race. One notable discrepancy, however, is that the best fit line (in solid blue) arcs above the 45-degree line in all cases. This owes to the fact that the mortgage application data shows higher first name concentration within each racial group than the voter file data. For example, the mortgage data shows that 2.6\% of White people are named Michael while our data indicates that only 1.7\% of White people are named Michael. The discrepancy may be driven by a complex interplay with the gender distribution of the datasets. The voter file data is slightly biased toward women, with 53.7\% of the registered individuals identifying as female vs. 50.5\% for the population at large \cite{census:2020}. Yet the mortgage application data appears to be biased to a greater degree, but in the opposite direction. Using the R package \texttt{gender} \cite{gender}, we are able to estimate gender probabilities for 97\% of the names in the mortgage dataset, from which we estimate that approximately 44.2\% of the individuals in the underlying dataset are female. Because there is greater concentration among male first names (that is, a higher proportion of men are given popular male first names than women are given popular female first names), the bias toward men in the mortgage data may induce greater name concentration overall \cite{census1990}. 

\section*{Usage Notes}

We provide a new dataset to estimate race-name distributions for first, middle, and last names, derived from publicly available voter files in six Southern states. These dictionaries, each comprising about one million names, are a rich and novel resource. The surname dictionary contains nearly ten times as many names as the the census surname list, the best public resource for estimating surname-race distributions. The first name dictionary contains more than 200 times as many names as the Tzioumis mortgage application dataset, which is the only first name data set currently available. The middle name dictionary is, to our knowledge, the first such dictionary of its kind. Having access to these dictionaries means that race imputation tasks can be conducted on a wider subset of the population, allowing public policy researchers to more credibly evaluate disparate racial impacts. 

While our analysis shows that the dictionaries are reasonably representative of national race distributions, they are not drawn from a random sample of the U.S. population. Registered voters must be over 18 and must be U.S. citizens, and they also typically differ from non-voters in terms of socioeconomic status. Moreover, our data is drawn from Southern states that have larger Black populations, but smaller Asian and Hispanic populations, than the country as a whole. Researchers should carefully consider the populations on which they would like to deploy these data. When considering surnames, it may often make sense to use our dictionary to supplement --- rather than supplant --- the census surname list, such that national race distributions are used when available. This is the approach taken in \cite{2205.06129}. 

The final usage note concerns the definition of the ``Other'' category. The racial categorization provided with these data is appropriate for many race imputation tasks, but may be problematic if researchers are interested in finer categorizations of individuals. In particular, these dictionaries may not be useful for identifying Native American communities, and hence should be used with caution in states with large Native American populations, such as Alaska and Oklahoma. 

\newpage

\section*{Acknowledgments}
We thank Bruce Willsie, CEO of L2, Inc., for providing us with the voter files we use in this paper.

\bibliography{my,imai,biblio}

\clearpage
\appendix

\section{Analysis of Name Discrepancies Between Census and Voter Files for ``Other'' Category}

Among the ten most common names such that $\mathbb{P}(\text{race} = \text{``Other''} \mid \text{name})$ differs by more than 25 percentage points between the census and the voter files, we find a notable commonality. At least six of the names -- ``Singh,'' ``Persaud,'' ``"Rampersad,'' ``Persad,'' ``Baksh,'' and ``Samaroo'' --- are recognizably of South Asian or Indo-Caribbean origin. Among these names, the Census typically categorizes more individuals as Asian, while the voter file data more typically categorizes these individuals as Other. This may explain the relatively low concordance seen between the datasets for this group only. The full list of names is provided in Table \ref{tab:discrepantNames}. 

\begin{table}[]
\begin{tabular}{l|rrrrr|rrrrr|r}
          & \multicolumn{5}{c}{Census Probabilities} & \multicolumn{5}{c}{Voter File Probabilities} &                                                                        \\ \toprule
Surname   & White & Black & Hispanic & Asian & Other & White  & Black  & Hispanic  & Asian  & Other & \begin{tabular}[c]{@{}l@{}}\# of Voter File \\Appearances\end{tabular} \\ \midrule
SINGH     & 4\%   & 4\%   & 3\%      & 83\%  & 6\%   & 9\%    & 5\%    & 3\%       & 50\%   & 33\%  & 5,945                                                                   \\
PERSAUD   & 10\%  & 21\%  & 5\%      & 52\%  & 13\%  & 7\%    & 8\%    & 6\%       & 35\%   & 45\%  & 1,770                                                                   \\
MOHAMED   & 20\%  & 66\%  & 2\%      & 8\%   & 4\%   & 14\%   & 42\%   & 3\%       & 12\%   & 30\%  & 1,523                                                                   \\
JOE       & 12\%  & 25\%  & 3\%      & 23\%  & 37\%  & 7\%    & 77\%   & 0\%       & 10\%   & 5\%   & 756                                                                    \\
RAMPERSAD & 6\%   & 43\%  & 4\%      & 28\%  & 18\%  & 11\%   & 14\%   & 3\%       & 28\%   & 44\%  & 349                                                                    \\
PERSAD    & 10\%  & 35\%  & 5\%      & 35\%  & 15\%  & 8\%    & 7\%    & 5\%       & 33\%   & 46\%  & 339                                                                    \\
BAKSH     & 9\%   & 26\%  & 5\%      & 48\%  & 12\%  & 6\%    & 9\%    & 3\%       & 29\%   & 52\%  & 315                                                                    \\
ALLY      & 27\%  & 23\%  & 4\%      & 40\%  & 6\%   & 23\%   & 16\%   & 5\%       & 21\%   & 35\%  & 282                                                                    \\
SAMAROO   & 8\%   & 32\%  & 4\%      & 44\%  & 12\%  & 8\%    & 12\%   & 4\%       & 29\%   & 48\%  & 269                                                                    \\
ALLI      & 24\%  & 35\%  & 5\%      & 30\%  & 7\%   & 15\%   & 20\%   & 7\%       & 23\%   & 35\%  & 266 \\ \bottomrule
\end{tabular}
\caption{\label{tab:discrepantNames} Ten most common names in the voter files such that the estimate of $\mathbb{P}(\text{race} = \text{``Other''} \mid \text{name})$ differs by more than 25 percentage points between the Census surname list and the voter file estimate. We observe that many of the names appear to be of South Asian or Indo-Caribbean origin.} 
\end{table}

\end{document}